\documentclass[twocolumn,aps,pra,showpacs]{revtex4}
\usepackage[T1]{fontenc}
\usepackage[latin1]{inputenc}
\usepackage{mathrsfs}
\usepackage{amsmath}
\usepackage{amssymb}
\usepackage{graphicx}
\usepackage{esint}

\makeatletter
\@ifundefined{textcolor}{}
{%
 \definecolor{BLACK}{gray}{0}
 \definecolor{WHITE}{gray}{1}
 \definecolor{RED}{rgb}{1,0,0}
 \definecolor{GREEN}{rgb}{0,1,0}
 \definecolor{BLUE}{rgb}{0,0,1}
 \definecolor{CYAN}{cmyk}{1,0,0,0}
 \definecolor{MAGENTA}{cmyk}{0,1,0,0}
 \definecolor{YELLOW}{cmyk}{0,0,1,0}
}

\makeatother

\begin{document}

\title{Soliton-induced Majorana fermions in a one-dimensional atomic topological
superfluid}

\author{Xia-Ji Liu$^{1}$ }

\email{xiajiliu@swin.edu.au}

\affiliation{$^{1}$Centre for Quantum and Optical Science, Swinburne University
of Technology, Melbourne 3122, Australia}

\date{\today}
\begin{abstract}
We theoretically investigate the behavior of dark solitons in a one-dimensional
spin-orbit coupled atomic Fermi gas in harmonic traps, by solving
self-consistently the Bogoliubov-de Gennes equations. The dark soliton
- to be created by phase-imprinting in future experiments - is characterized
by a real order parameter, which changes sign at a point node and
hosts localized Andreev bound states near the node. By considering
both cases of a single soliton and of multiple solitons, we find that
the energy of these bound states decreases to zero, when the system
is tuned to enter the topological superfluid phase by increasing an
external Zeeman field. As a result, two Majorana fermions emerge in
the vicinity of each soliton, in addition to the well-known Majorana
fermions at the trap edges associated with the nontrivial topology
of the superfluid. We propose that the soliton-induced Majorana fermions
can be directly observed by using spatially-resolved radio-frequency
spectroscopy or indirectly probed by measuring the density profile
at the point node. For the latter, the deep minimum in the density
profile will disappear due to the occupation of the soliton-induced
zero-energy Majorana fermion modes. Our prediction could be tested
in a resonantly-interacting spin-orbit coupled $^{40}$K Fermi gas
confined in a two-dimensional optical lattice. 
\end{abstract}

\pacs{03.75.Ss, 71.10.Pm, 03.65.Vf, 03.67.Lx}

\maketitle

\section{Introduction}

Solitonic excitations in quantum superfluids are of significant importance.
They behave like quantum mechanical matter waves and maintain their
shape during propagation. As highly nonlinear and localized topological
excitations, solitons provide a very sensitive probe of the fundamental
coherence properties of the underlying superfluid state in which they
propagate. In the ultracold matter of weakly interacting Bose-Einstein
condensates (BECs), where the phase coherence is associated with long-range
order in the one-body density matrix, solitons have been investigated
extensively over the past decade, both theoretically and experimentally
\cite{Frantzeskakis2010}. In a series of ground-breaking experiments
\cite{Burger1999,Denschlag2000}, dark solitons - appearing as a suppression
in the density profile - have been created via phase-imprinting. Their
theoretical description is provided by the nonlinear Gross-Pitaevskii
equation \cite{Frantzeskakis2010}. 

In strongly interacting fermionic superfluids, solitions are more
interesting, as the phase coherence between the underlying bosonic
entities (i.e., Cooper pairs), characterized by long-range order in
the two-body density matrix, is more subtle to understand and describe
\cite{Dziarmaga2005,Antezza2007,Liao2011,Scott2011,Spuntarelli2011,Cetoli2013}.
In addition, fermionic bound states may be induced near solitonic
defects \cite{Antezza2007}, analogous to the famous Andreev bound
states inside vortex cores \cite{Caroli1964}. In this respect, the
recent experimental realization of dark solitons in strongly-interacting
atomic $^{6}$Li Fermi gases at the crossover from BECs to Bardeen-Cooper-Schrieffer
(BCS) superfluids is of great interest \cite{Yefsah2013}. A fermionic
soliton was nucleated in a controlled way by using the phase imprinting
method in a cigar-shaped Fermi cloud and was observed as a reduced
density slit running through the middle of the cloud. The soliton
exhibited the expected oscillation when it moved from one end of the
trap to the other. However, the rate of oscillation was much slower
than that predicted from time-dependent mean-field calculations \cite{Liao2011,Scott2011,Spuntarelli2011,Cetoli2013}.
This discrepancy is now solved by a refined measurement \cite{Ku2014}.
Due to the intrinsic snake instability in three dimensions, the observed
defect is actually the decay product of soliton. In the constrained
geometry of the cloud, it is a single straight vortex line and therefore
is better named as a solitonic vortex \cite{Ku2014}.

In this work, we consider the observation of dark solitons in an even
more intriguing situation - topological fermionic superfluids \cite{Qi2011}.
Our research is motivated by the recent opened perspective that atomic
topological superfluids might be realized very soon in cold-atom laboratories
with spin-orbit coupled Fermi gases of $^{40}$K or $^{6}$Li atoms
\cite{Wang2012,Cheuk2012,Williams2013}. Topological superfluids are
novel states of matter, which have attracted considerable attention
because of their non-trivial topological properties and their ability
to host exotic quasi-particles known as Majorana fermions - particles
that are their own anti-particles \cite{Majorana1937,Wilczek2009}.
It is therefore of great interest to ask: will there be any interesting
features exhibited by a superfluid when its topological order and
a solitonic defect come into play?

To address this problem, we theoretically investigate the existence
of dark solitons in a one-dimensional (1D) spin-orbit coupled atomic
Fermi gas in a harmonic trap under an external Zeeman field. A dark
soliton is characterized by a phase jump of $180^{\circ}$ at a point
node at which the order parameter changes sign and crosses zero. As
the number of dark solitons can be controlled by changing the number
of phase jumps, we consider both cases of a single soliton and multiple
solitons (i.e., a soliton train). In the topological superfluid phase,
we find that each dark soliton is able to host two Majorana fermions,
well localized around nodal point of the soliton. Potentially, this
may provide an ideal scenario to create and move Majorana fermions
towards realistic applications, via the control of phase-imprinting.
We propose that experimentally the existence of dark solitons may
be probed by using radio-frequency spectroscopy for the local density
of state or absorption imaging for the density profile.

It should be noted that a dark soliton in one dimension behaves very
similarly to a vortex in two dimensions. The latter is also a topological
defect that can host a Majorana fermion in the vortex core and exhibit
it in the local density of state and density profile \cite{Liu2012a}.
A vortex lattice in topological superfluids has been proposed to be
an appealing platform to perform topological quantum information processing
and quantum computation \cite{Nayak2008,Ivanov2001}. In principle,
a soliton train would achieve a similar goal. A detailed discussion
of this possibility will be addressed in a future publication.

We also note that recently Xu \textit{et al.} investigated the properties
of a single soliton in a 1D spin-orbit coupled Fermi gas \cite{Xu2014}.
These authors considered a different set of parameters, with which
the Fermi cloud enters a partial topological superfluid phase by increasing
the external Zeeman field. Our results are in qualitative agreement
with theirs when there are overlaps.

Our paper is arranged as follows. In the next section (Sec. II), we
briefly introduce the model Hamiltonian and explain how to solve it
in the mean-field picture of the Bogoliubov-de Gennes (BdG) equations,
and then specify the parameter space (i.e., phase diagram) for our
numerical results. In Sec. III, we study the properties of a single
soliton and show the emergence of zero-energy Majorana fermions when
the topological regime is approached. The wavefunctions of the Majorana
fermions and their manifestations in the local density of state and
density profile are discussed in detail. In Sec. IV, the cases of
two and more solitons are considered. Finally, Sec. V is devoted to
summaries and conclusions.

\section{Model Hamiltonian and mean-field theory}

We consider a spin-1/2 $^{40}$K Fermi gas of $N$ atoms with spin-orbit
coupling confined in a 1D harmonic trap \cite{Liu2012b,Wei2012,Liu2013,Jiang2014}.
This system could be realized at Shanxi University, by adding a very
deep 2D optical lattice (in the transverse $y$-$z$ plane) to a spin-orbit
coupled 3D Fermi gas formed by two counter-propagating Raman laser
beams (along the $x$-direction) \cite{Wang2012}. It can be described
by a single-channel model Hamiltonian $H=\int dx[\mathcal{\mathscr{H}}_{0}+\mathcal{\mathscr{H}}_{int}${]},
where 
\begin{equation}
\mathscr{H}_{0}=\sum_{\sigma=\uparrow,\downarrow}\Psi_{\sigma}^{\dagger}\mathcal{H}_{S}\Psi_{\sigma}-\frac{\Omega_{R}}{2}\left[\Psi_{\uparrow}^{\dagger}e^{i2k_{R}x}\Psi_{\downarrow}+\text{H.c.}\right]\label{bareHami1}
\end{equation}
is the single-particle part and 
\begin{equation}
\mathscr{H}_{int}=g_{1D}\Psi_{\uparrow}^{\dagger}\left(x\right)\Psi_{\downarrow}^{\dagger}\left(x\right)\Psi_{\downarrow}\left(x\right)\Psi_{\uparrow}\left(x\right)
\end{equation}
is the part describing the contact interaction between the two spin
states. Here, $\sigma=\uparrow,\downarrow$ is the pseudo-spin denoting
the two hyperfine states and $\Psi_{\sigma}^{\dagger}\left(x\right)$
is the fermionic field operator that creates an atom with mass $m$
in the spin state $\sigma$. The second term in $\mathscr{H}_{0}$
describes a synthetic spin-orbit coupling, where $\Omega_{R}$ and
$k_{R}$ are the Rabi frequency and the wavevector of the laser beams,
respectively. Due to the counter-propagating configuration, the momentum
transferred during the two-photon Raman process is $2\hbar k_{R}$.
The Hamiltonian 
\begin{equation}
\mathcal{H}_{S}=-\frac{\hbar^{2}}{2m}\frac{\partial^{2}}{\partial x^{2}}-\mu+V_{T}(x),
\end{equation}
with $\mu$ being the chemical potential and $V_{T}\left(x\right)\equiv m\omega^{2}x^{2}/2$
the 1D harmonic trapping potential with an oscillation frequency $\omega$.
The motion of atoms in the $y$-$z$ plane is frozen due to the tight
2D optical lattice (i.e., the transverse trapping frequency $\omega_{\perp}\sim N\omega\gg\omega$).
In this quasi-1D geometry, it is known that the low-energy scattering
properties of atoms can be well described using a contact potential
$g_{1D}\delta(x)$, where the 1D effective coupling constant $g_{1D}<0$
can be expressed through the 3D scattering length $a_{3D}$ \cite{Bergeman2003},

\begin{equation}
g_{1D}=\frac{2\hbar^{2}a_{3D}}{ma_{\perp}^{2}}\frac{1}{\left(1-{\cal A}a_{3D}/a_{\perp}\right)},
\end{equation}
where $a_{\perp}\equiv\sqrt{\hbar/(m\omega_{\perp})}$ and the constant
${\cal A}=-\zeta(1/2)/\sqrt{2}\simeq1.0326$. The interatomic interaction
of our trapped system can be conveniently parameterized by a dimensionless
interaction parameter \cite{Hu2007a,Liu2007,Liu2008} 
\begin{equation}
\gamma\equiv-\frac{mg_{1D}}{\hbar^{2}n_{0}},
\end{equation}
which is basically the ratio between the mean-field interaction energy
and the kinetic energy. Here $n_{0}=(2/\pi)\sqrt{Nm\omega/\hbar}$
is the total atomic density of a non-interacting Fermi gas at the
trap center within the local density approximation. Experimentally,
in the vicinity of Feshbach resonances (i.e., $a_{3D}\rightarrow\pm\infty$),
the typical value of the interaction parameter $\gamma$ is about
$5$ \cite{Liao2010}.

To solve the single-channel model Hamiltonian $H$, it is useful to
apply a local gauge transformation \cite{Liu2012b,Wei2012,Jiang2014},
\begin{eqnarray}
\Psi_{\uparrow}\left(x\right) & = & e^{+ik_{R}x}\frac{1}{\sqrt{2}}\left[\psi_{\uparrow}\left(x\right)-i\psi_{\downarrow}\left(x\right)\right],\\
\Psi_{\downarrow}\left(x\right) & = & e^{-ik_{R}x}\frac{1}{\sqrt{2}}\left[\psi_{\uparrow}\left(x\right)+i\psi_{\downarrow}\left(x\right)\right].
\end{eqnarray}
Using the new field operators $\psi_{\uparrow}\left(x\right)$ and
$\psi_{\downarrow}\left(x\right)$, we can recast the single-particle
Hamiltonian into the form, 
\begin{eqnarray}
\mathscr{H}_{0} & = & \left[\psi_{\uparrow}^{\dagger}\left(x\right),\psi_{\downarrow}^{\dagger}\left(x\right)\right]{\cal H}_{0}\left[\begin{array}{c}
\psi_{\uparrow}\left(x\right)\\
\psi_{\downarrow}\left(x\right)
\end{array}\right],\\
{\cal \mathcal{H}}_{0} & = & -\frac{\hbar^{2}}{2m}\frac{\partial^{2}}{\partial x^{2}}+V_{T}\left(x\right)-\mu-h\sigma_{z}+\lambda\hat{k}_{x}\sigma_{y},
\end{eqnarray}
where we have redefined the chemical potential $\mu\rightarrow\mu-\hbar^{2}k_{R}^{2}/(2m)$
to absorb a constant energy shift and have introduced the momentum
operator $\hat{k}_{x}\equiv-i\partial/\partial x$, the spin-orbit
coupling constant $\lambda\equiv\hbar^{2}k_{R}/m$ and an effective
Zeeman field $h\equiv\Omega_{R}/2$. Furthermore, $\sigma_{y}$ and
$\sigma_{z}$ are Pauli matrices. The form of the interaction Hamiltonian
is invariant under the local gauge transformation, i.e., 
\begin{equation}
\mathscr{H}_{int}=g_{1D}\psi_{\uparrow}^{\dagger}\left(x\right)\psi_{\downarrow}^{\dagger}\left(x\right)\psi_{\downarrow}\left(x\right)\psi_{\uparrow}\left(x\right).
\end{equation}
The operator for the total atomic density $\hat{n}(x)\equiv\sum_{\sigma}\Psi_{\sigma}^{\dagger}\left(x\right)\Psi_{\sigma}\left(x\right)=\sum_{\sigma}\psi_{\sigma}^{\dagger}\left(x\right)\psi_{\sigma}\left(x\right)$
is also invariant. However, the form of the density operator for each
spin component changes \cite{Wei2012}.

\subsection{Bogoliubov-de Gennes equations}

We solve the single-channel model Hamiltonian within the standard
mean-field framework. By introducing an order parameter $\Delta\left(x\right)\equiv-g_{1D}\left\langle \psi_{\downarrow}\left(x\right)\psi_{\uparrow}\left(x\right)\right\rangle $,
the interaction Hamiltonian can be approximated as, 
\begin{equation}
\mathscr{H}_{int}\simeq-\left[\Delta\left(x\right)\psi_{\uparrow}^{\dagger}\left(x\right)\psi_{\downarrow}^{\dagger}\left(x\right)+\text{H.c.}\right]-\frac{\left|\Delta\left(x\right)\right|^{2}}{g_{1D}}.
\end{equation}
It is then convenient to use a Nambu spinor ${\bf \mathbf{\boldsymbol{\psi}}}(x)\equiv[\psi_{\uparrow}\left(x\right),\psi_{\downarrow}\left(x\right),\psi_{\uparrow}^{\dagger}\left(x\right),\psi_{\downarrow}^{\dagger}\left(x\right)]^{T}$
and rewrite the model Hamiltonian in a compact form, 
\begin{equation}
H_{mf}=\frac{1}{2}\int dx{\bf \boldsymbol{\psi}}^{\dagger}{\cal H}_{BdG}{\bf \boldsymbol{\psi}}+\text{Tr}{\cal H}_{S}-\int dx\frac{\left|\Delta\left(x\right)\right|^{2}}{g_{1D}},
\end{equation}
where

\begin{equation}
{\cal H}_{BdG}\equiv\left[\begin{array}{cccc}
{\cal H}_{S}-h & -\lambda\partial/\partial x & 0 & -\Delta(x)\\
\lambda\partial/\partial x & {\cal H}_{S}+h & \Delta(x) & 0\\
0 & \Delta^{*}(x) & -{\cal H}_{S}+h & \lambda\partial/\partial x\\
-\Delta^{*}(x) & 0 & -\lambda\partial/\partial x & -{\cal H}_{S}-h
\end{array}\right]\label{BdG_Hami}
\end{equation}
and the term Tr${\cal H}_{S}$ comes from the anti-commutativity of
Fermi field operators. The mean-field model Hamiltonian can then be
diagonalized as
\begin{equation}
{\cal H}_{BdG}\Phi_{\eta}\left(x\right)=E_{\eta}\Phi_{\eta}\left(x\right),\label{bdgeq}
\end{equation}
where $\Phi_{\eta}(x)\equiv[u_{\uparrow\eta}\left(x\right),u_{\downarrow\eta}\left(x\right),v_{\uparrow\eta}\left(x\right),v_{\downarrow\eta}\left(x\right)]^{T}$
and $E_{\eta}$ are respectively the wave-function and the energy
of Bogoliubov quasiparticles, indexed by an integer subscript $\eta=1,2,3,\cdots$.
The BdG Hamiltonian Eq. (\ref{BdG_Hami}) includes the order parameter
$\Delta\left(x\right)$ that should be determined self-consistently:
\begin{equation}
\Delta(x)=-\frac{g_{1D}}{2}\sum_{\eta}\left[u_{\uparrow\eta}v_{\downarrow\eta}^{*}f\left(E_{\eta}\right)+u_{\downarrow\eta}v_{\uparrow\eta}^{*}f\left(-E_{\eta}\right)\right],\label{gapeq}
\end{equation}
where $f\left(E\right)\equiv1/[e^{E/k_{B}T}+1]$ is the Fermi distribution
function at temperature $T$. The chemical potential $\mu$ can be
determined using the number equation, $N=\int dxn\left(x\right)$,
where the total atomic density is given by 
\begin{equation}
n\left(x\right)=\frac{1}{2}\sum_{\sigma\eta}\left[\left|u_{\sigma\eta}\right|^{2}f\left(E_{\eta}\right)+\left|v_{\sigma\eta}\right|^{2}f\left(-E_{\eta}\right)\right].\label{numeq}
\end{equation}
It is worth noting that the use of Nambu spinors double the Hilbert
space of the system. As a consequence, we always have an intrinsic
particle-hole symmetry in the Bogoliubov solutions. That is, for any
``particle'' solution with the wave-function $\Phi_{\eta}^{(p)}(x)=[u_{\uparrow\eta},u_{\downarrow\eta},v_{\uparrow\eta},v_{\downarrow\eta}]^{T}$
and energy $E_{\eta}^{(p)}\geq0$, there is another partner ``hole''
solution with $\Phi_{\eta}^{(h)}(x)=[v_{\uparrow\eta}^{*},v_{\downarrow\eta}^{*},u_{\uparrow\eta}^{*},u_{\downarrow\eta}^{*}]^{T}$
and $E_{\eta}^{(h)}=-E_{\eta}^{(p)}\leq0$. These two solutions actually
correspond to the same physical quantum state. To avoid double counting,
an extra factor of 1/2 appears in the expression for the order parameter
Eq. (\ref{gapeq}) and the total atomic density Eq. (\ref{numeq}).

The Bogoliubov equation Eq. (\ref{bdgeq}) can be solved iteratively
with Eqs. (\ref{gapeq}) and (\ref{numeq}) by using a basis expansion
method, together with a hybrid strategy that takes care of the high-lying
energy states \cite{Liu2012b,Liu2013,Liu2007,Liu2008}. Once we have
the solution, we calculate straightforwardly the local density of
states,
\begin{equation}
\rho\left(x,\omega\right)=\frac{1}{2}\sum_{\sigma\eta}\left[\left|u_{\sigma\eta}\right|^{2}\delta\left(\omega-E_{\eta}\right)\right.+\left.\left|v_{\sigma\eta}\right|^{2}\delta\left(\omega+E_{\eta}\right)\right].
\end{equation}

\begin{figure}
\begin{centering}
\includegraphics[clip,width=0.48\textwidth]{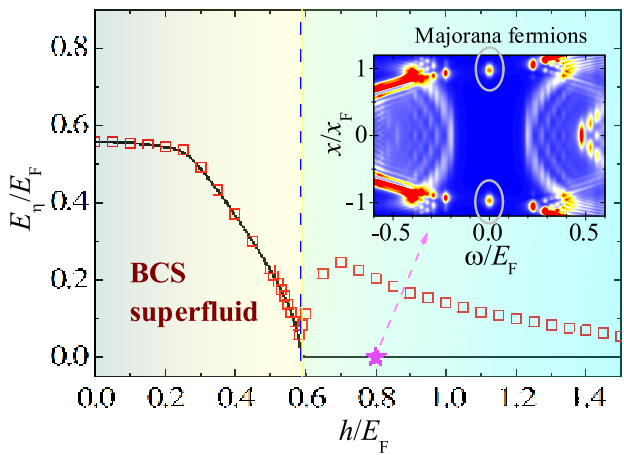} 
\par\end{centering}

\protect\caption{(color online) Zero-temperature phase diagram without solitons at
$\gamma=2.2$ and $\lambda=1.5E_{F}/k_{F}$, determined from the behavior
of the two lowest-energy particle solutions (with $E_{\eta}\geq0$)
in the Bogoliubov quasiparticle spectrum (shown by the solid line
and empty squares, respectively). The topological phase transition
occurs at about $h\simeq0.59E_{F}$. As a result, two zero-energy
Majorana fermions emerge at the trap edges, as shown in the inset,
plotting the local density of states for the case $h=0.8E_{F}$. They
are highlighted by two grey circles.}

\label{fig1} 
\end{figure}

We note that in low dimensions phase fluctuations are generally enhanced
due to the reduced dimenionality. As a result, in one dimension a
true long-range order (i.e., characterized by an order parameter)
is ruled out by the well-known Mermin\textendash Wagner\textendash Hohenberg
theorem \cite{Mermin1966,Hohenberg1967}. Nevertheless, at zero temperature
the 1D pair correlation shows a slow power-law decay, much slower
than an exponential decay expected for a normal state, as predicted
by using bosonization or exact Bethe ansatz approach \cite{Guan2013}.
Thus, we anticipate that the assumption of an order parameter and
the use of mean-field BdG equations may still provide a useful approximation.
Indeed, without spin-orbit coupling, we have checked that the mean-field
approach leads to reasonably accurate equation of state for a weak-coupling
uniform Fermi gas, compared with the exact Gaudin-Yang solution \cite{Liu2007}.
In harmonic traps, the mean-field theory also predicts very similar
density profiles as the Bethe ansatz approach (within the local density
approximation) \cite{Liu2007}. Therefore, in the presence of spin-orbit
coupling, we assume that our mean-field treatment may qualitatively
capture the essential physics of fermionic solitons.

\subsection{Solitonic order parameter}

A stationary dark soliton is characterized by a $\pi$-phase jump.
In this case, the order parameter may be chosen to be \emph{real}
and a stationary soliton at the point node $x_{1}$ may take the following
form for the order parameter:
\begin{equation}
\Delta\left(x\right)=\left|\Delta\left(x\right)\right|\exp\left[i\pi\Theta\left(x-x_{1}\right)\right],\label{eq:gap1soliton}
\end{equation}
where $\Theta(x)$ is the Heaviside step function. Similarly, the
order parameter of a soliton train with point nodes at $x_{i}$ ($i=1,2,3,\cdots$)
is given by,
\begin{equation}
\Delta\left(x\right)=\left|\Delta\left(x\right)\right|\exp\left[i\pi\sum_{i}\Theta\left(x-x_{i}\right)\right].\label{eq:gapMsoliton}
\end{equation}
In numerical calculations, we calculate the order parameter through
Eq. (\ref{gapeq}) and then update it by phase-imprinting the $\pi$-phase
jumps with the use of Eq. (\ref{eq:gap1soliton}) or Eq. (\ref{eq:gapMsoliton}).
The solitonic order parameter is obtained self-consistently, after
a number of iterations up to convergence.

\subsection{Phase diagram without solitons}

The solution of the BdG equations without solitons has been discussed
in detail in earlier works \cite{Liu2012b,Liu2013,Jiang2014}. In
Fig. \ref{fig1}, we present the typical phase diagram at zero temperature
for a spin-orbit coupled Fermi gas with $N=60$ atoms at the interaction
strength $\gamma=2.2$ and spin-orbit coupling strength $\lambda=1.5E_{F}/k_{F}$.
Here, we use the Thomas-Fermi energy and wavevector of a non-interacting
trapped Fermi gas, $E_{F}=(N/2)\hbar\omega$ and $k_{F}=\sqrt{2mE_{F}}$,
as the units of energy and wavevector, respectively. The units of
length are given by the Thomas-Fermi radius, $x_{F}=\sqrt{N\hbar/(m\omega)}$.
Throughout this work, we focus on the case of zero temperature and
shall use the same total number of atoms ($N=60$), interaction parameter
($\gamma=2.2$) and spin-orbit coupling strength ($\lambda=1.5E_{F}/k_{F}$),
unless otherwise specified. For the basis expansion, we have considered
540 single-particle eigenstates of the harmonic oscillator $\mathcal{H}_{S}$
and have taken a cut-off energy $E_{c}=15E_{F}=450\hbar\omega$, above
which we use the local density approximation \cite{Liu2012b,Liu2013}.

Fig. \ref{fig1} shows the energies of the two lowest-energy \emph{particle}
solutions with $E_{\eta}\geq0$, plotted respectively by using the
solid line and empty squares. A topological phase transition occurs
at a critical effective Zeeman field $h_{c}\simeq0.59E_{F}$ \cite{Lutchyn2010,Oreg2010,footnote1},
as revealed by $\min\{E_{\eta}\}$ (solid line). At a small Zeeman
field $h<h_{c}$, the Fermi gas is a standard BCS superfluid, with
a fully gapped quasiparticle energy spectrum (i.e., $\min\{E_{\eta}\}>0$).
Above the threshold, $h>h_{c}$, the quasiparticle energy spectrum
is again gapped in the \emph{bulk}, as seen from the empty squares.
However, gapless edge excitations - Majorana fermions - emerge at
the trap edges. This is particularly evident when we plot the local
density of states in the inset. Two (nearly) zero-energy Majorana
fermions - well localized at the trap edges as highlighted by the
two grey circles - are clearly visible.

It is useful to note that Majorana fermions may acquire an exponentially
small energy due to the finite size of the harmonic trap \cite{Potter2010}.
The typical spatial extension of the localized Majorana wave-function
$\xi_{M}$ is at the order of the coherence length $\xi_{c}=\hbar v_{F}/\Delta$,
where $v_{F}$ and $\Delta$ are the unperturbed local Fermi velocity
and pairing gap at the trap edge \cite{Potter2010}. For the two Majorana
fermions shown in the inset of Fig. \ref{fig1}, we estimate that
$\xi_{M}\sim0.1x_{F}$ (see also Fig. \ref{fig4}). Therefore, the
exponentially small overlap between two Majorana fermion wave-functions
$\Phi_{L}$ and $\Phi_{R}$, i.e., $\mathcal{O}=\left\langle \Phi_{L}|\Phi_{R}\right\rangle \sim\exp(-L/\xi_{M})$,
where $L\simeq2x_{F}$ is the distance between two Majorana fermions,
leads to an exponentially small energy (splitting) of Majorana fermions,
\begin{equation}
\frac{E}{E_{F}}\sim\mathcal{O}\sim\exp\left(-\frac{L}{\xi_{M}}\right)\sim10^{-9}.\label{eq:energysplitting}
\end{equation}
This is consistent with our numerical finding that the energy of Majorana
fermions $E\sim10^{-10}E_{F}$.

\section{Single soliton}

Here we consider the behavior of a single dark soliton. The case of
multiple dark solitons will be discussed in the next section.

\begin{figure}
\begin{centering}
\includegraphics[clip,width=0.48\textwidth]{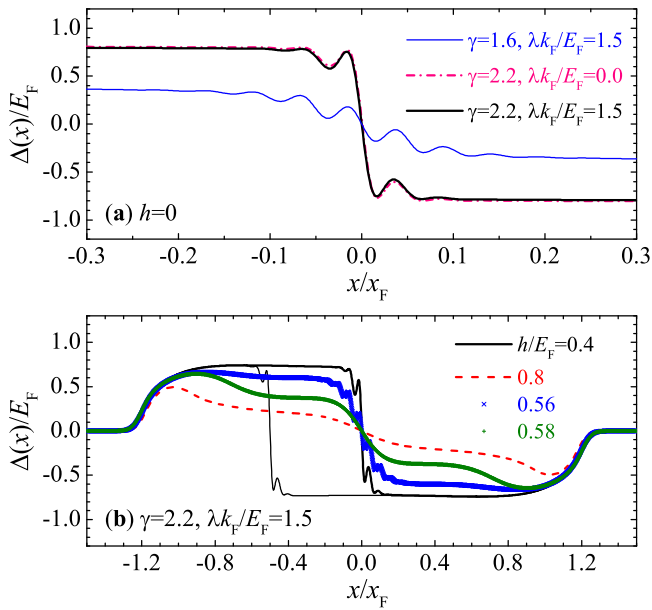} 
\par\end{centering}

\protect\caption{(color online) (a) The solitonic pairing gap $\Delta(x)$ at zero
Zeeman field $h=0$ and at different interaction strengths $\gamma$
and spin-orbit coupling strengths $\lambda$, when a $\pi$-phase
jump is imprinted at $x_{1}=0$. (b) The solitonic $\Delta(x)$ at
nonzero Zeeman fields {[}$h/E_{F}=0.4$ (thick solid line), $0.56$
(crosses), $0.58$ (daggers) or $0.8$ (dashed line){]} and at $\gamma=2.2$
and $\lambda=1.5E_{F}/k_{F}$. For the case of $h=0.4E_{F}$, we also
consider a soliton at $x_{1}=-0.5x_{F}$ and plot the result with
a thin solid line.}

\label{fig2} 
\end{figure}

\subsection{Order parameters}

In Fig. \ref{fig2}, we report the pairing gap profile in the presence
of a single dark soliton, at zero Zeeman field (a) or finite Zeeman
fields (b). The order parameter crosses zero at the position of the
soliton $x_{1}$ and hence creates a point node. At small Zeeman fields,
it exhibits two length scales around the point node \cite{Antezza2007}:
a fast oscillation with length scale of $k_{F}^{-1}$, and a slower
healing with length scale $\xi_{c}=\hbar v_{F}/\Delta$. Here, $v_{F}$
and $\Delta$ are the unperturbed local Fermi velocity and pairing
gap at the point node $x_{1}$, respectively. The former length scale
is essentially independent of the interaction parameter and the spin-orbit
coupling strength. Thus, as in the case of a vortex in 2D Fermi gases
\cite{Hu2007b}, we may safely identify the oscillation as the Friedel
oscillation. For the coherence length, we find that $\xi_{c}\simeq3k_{F}^{-1}$
at $\gamma=2.2$. It increases with decreasing interaction parameter,
as expected.

For a spin-orbit coupled Fermi gas, it is interesting to see (Fig.
\ref{fig2}(b)) that the Friedel oscillation suddenly ceases to exist
when the Zeeman field is above a threshold, $h_{c}\simeq0.57E_{F}$.
Actually, this point corresponds to the topological phase transition
in the presence of a single dark soliton, which we shall now discuss
in greater detail.

\begin{figure*}[t]
\begin{centering}
\includegraphics[clip,width=0.98\textwidth]{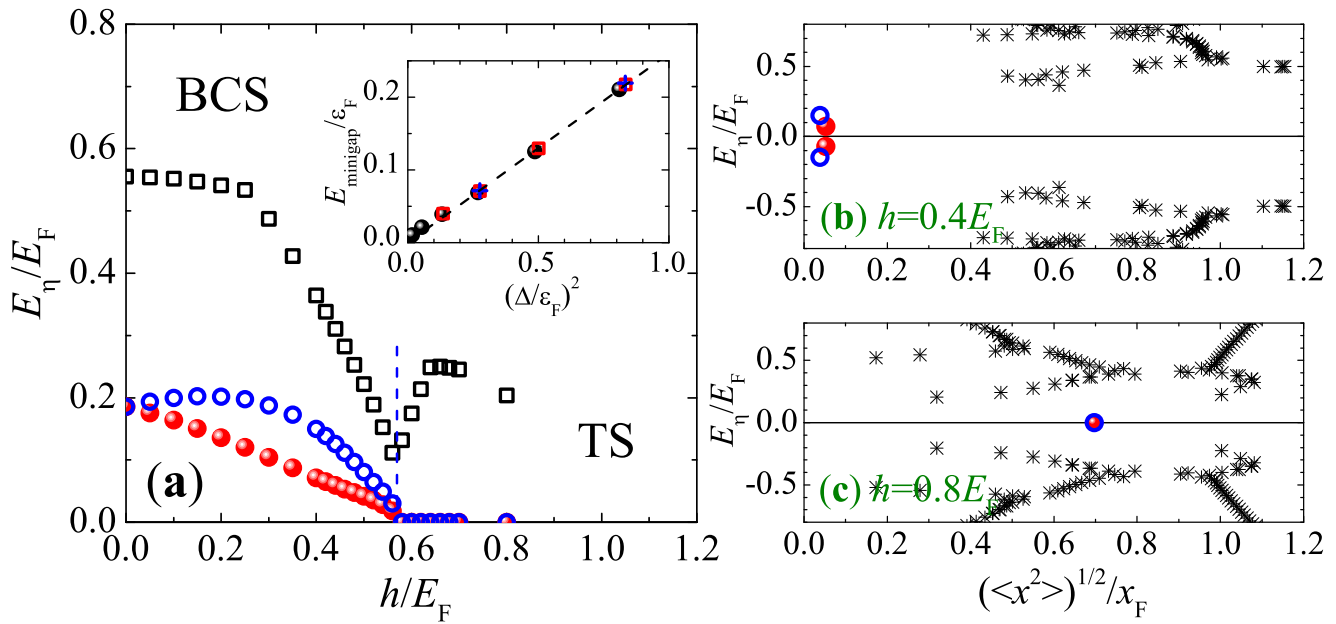} 
\par\end{centering}

\protect\caption{(color online) (a) The energies of the three lowest-energy particle
solutions with $E_{\eta}\geq0$, as a function of the Zeeman field
in the case of a single dark soliton at $x_{1}=0$. The solid and
empty circles show Andreev bound states localized near the soliton,
while the empty squares correspond to the lowest-energy particle state
in the bulk. The vertical dashed line separates the BCS and topological
superfluid phases. The inset shows the minigap at $h=0$ and at different
interaction parameters $\gamma$ and spin-orbit coupling strengths
{[}$\lambda k_{F}/E_{F}=1.5$ (circles), $1.0$ (squares) and $0.0$
(daggers){]}, as a function of the square of the local pairing gap,
$\Delta^{2}/\epsilon_{F}^{2}$. The dashed line is the linear fit
$E_{\textrm{minigap}}=0.26\Delta^{2}/\epsilon_{F}$. (b) and (c) The
spatial distribution of Bogoliubov quasiparticle energy spectrum at
$h=0.4E_{F}$ and $h=0.8E_{F}$, respectively. The Andreev bound states
near the soliton are highlighted by empty and solid circles. Here,
we approximately characterize the location of a quasiparticle by using
its wave-function: $\sqrt{\left\langle x^{2}\right\rangle }=\{\int dx\: x^{2}\sum_{\sigma}[u_{\sigma}^{2}\left(x\right)+\nu_{\sigma}^{2}\left(x\right)]\}^{1/2}$,
if the wave-function is well-localized at a certain point. Otherwise,
$\sqrt{\left\langle x^{2}\right\rangle }$ characterizes the width
of the wave-function.}

\label{fig3} 
\end{figure*}

\subsection{Andreev bound states and Majorana fermions}

In the presence of a single dark soliton, the topological phase transition
point can also be determined from the Bogoliubov quasiparticle spectrum.
In Fig. \ref{fig3}(a), we plot the energies of the three lowest-energy
\emph{particle} solutions as a function of the Zeeman field, while
in Figs. \ref{fig3}(b) and \ref{fig3}(c), we show the characteristic
spectrum before and after the topological transition. The transition
is associated with the closing and reopening of the energy gap in
the bulk \cite{Qi2011,Liu2012a,Liu2012b,Wei2012,Lutchyn2010,Oreg2010},
see, for example, the empty squares in Fig. \ref{fig3}(a) for the
lowest-energy bulk state in the particle branch. Thus, we determine
that the transition occurs at the critical field $h_{c}\simeq0.57E_{F}$,
which is slightly smaller than the threshold $h_{c}\simeq0.59E_{F}$
in the absence of the dark soliton (see Fig. \ref{fig1}).

In the topologically-trivial BCS superfluid phase, we find two Andreev-like
bound states that reside near the point node of the soliton. Their
existence is easy to understand from the density dip of the solitonic
order parameter, which basically creates an effective confinement
potential of length scale $\xi_{c}$ for quasiparticles. As a result,
localized states develop, with a characteristic energy of the order
of $\hbar^{2}/(m\xi_{c}^{2})=\Delta^{2}/(2\epsilon_{F})$, where $\epsilon_{F}$
is the local Fermi energy. These are reminiscent of the well-known
Caroli-de Gennes-Matricon states in a vortex core in 2D Fermi gases
\cite{Caroli1964}. 

\begin{figure}
\begin{centering}
\includegraphics[clip,width=0.48\textwidth]{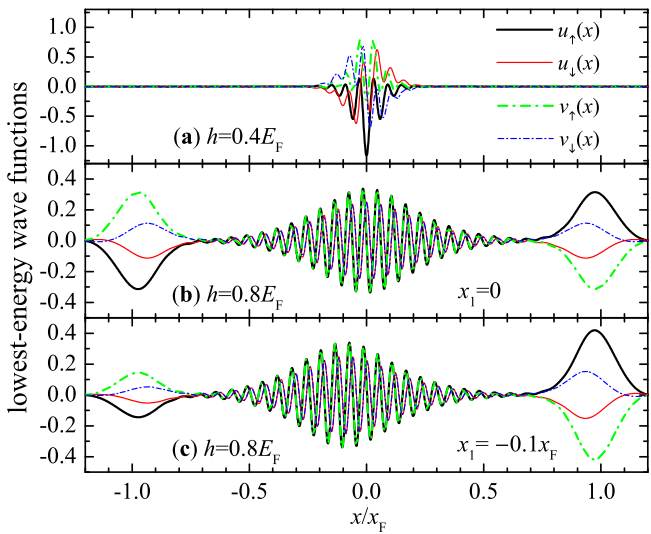} 
\par\end{centering}

\protect\caption{(color online) The wavefunctions of the lowest-energy Andreev bound
states in the BCS superfluid phase at $h=0.4E_{F}$ (a) or in the
topologically-nontrivial superfluid phase at $h=0.8E_{F}$ (b and
c). In the latter case, the wavefunctions strongly interfere with
that of Majorana fermions at the trap edges. The superposition persists
when we place the dark soliton in a less symmetric position $x_{1}=-0.1x_{F}$. }

\label{fig4} 
\end{figure}

At zero Zeeman field, the two Andreev bound states are degenerate
in energy, which is nonzero and is called the \emph{minigap} in the
context of superconductors in solid state physics. In the weakly-interacting
limit, as shown in the inset of Fig. \ref{fig3}(a), we have checked
that, to a very good approximation,
\begin{equation}
E_{\textrm{minigap}}\simeq0.26\frac{\Delta^{2}}{\epsilon_{F}}.
\end{equation}
With increasing Zeeman field, the energies of the two Andreev bound
states initially split: one increases and the other decreases. However,
when $h>0.2E_{F}$, both energies gradually decrease to zero as $h$
nears the topological transition. A typical energy spectrum before
the transition at $h=0.4E_{F}$ is shown in Fig. \ref{fig3}(b), plotted
as a function of the approximate location of each quasiparticle state,
\begin{equation}
\sqrt{\left\langle x^{2}\right\rangle }=\left[\int dx\: x^{2}\sum_{\sigma}\left(u_{\sigma}^{2}\left(x\right)+\nu_{\sigma}^{2}\left(x\right)\right)\right]^{1/2}.
\end{equation}
The two Andreev bound states, represented by the empty and solid circles
\cite{footnote3}, are clearly seen near the point node of the soliton
at $x_{1}=0$. The highly-localized wavefunctions of the lowest-energy
bound state, corresponding to the solid circle, are plotted in Fig.
\ref{fig4}(a).

Approaching the critical Zeeman field $h_{c}\simeq0.57E_{F}$, the
energies of the two Andreev bound states tend to zero. The energies
of some bulk states also vanish. The interference of these nearly
zero-energy modes leads to a huge reconstruction of the quasiparticle
spectrum across the topological transition. Immediately after the
transition, with the reopening of the energy gap in the bulk, we observe
that one of the Andreev bound states merges with bulk states and therefore
can no longer be identified as a localized state. At the same time,
two zero-energy edge states appear at the trap edges. As a result,
we find four Majorana fermions when the system is in the topological
phase: two at the edges and the other two near the soliton \cite{footnote3}.
Physically, the number of Majorana fermions may be understood from
the fact that a single dark soliton effectively splits the Fermi gas
into two, each of which could host two Majorana fermions. In the case
of multiple dark solitons, we therefore anticipate that the total
number of Majorana fermions would be $2n+2$, where $n$ is the number
of solitons. This expectation is confirmed by our numerical calculations.

In Figs. \ref{fig4}(b) and \ref{fig4}(c), we examine the wave functions
of the four Majorana fermions. Since all four states are degenerate
with zero energy, they may be mutually-superposed. The interference
leads to very similar wavefunctions, so in the figure, only one of
the four is plotted. The bond and anti-bond superpositions of the
well-localized Majorana wavefunctions, presumably \emph{one} at the
point node of the soliton and the other \emph{two} at the edges, are
fairly clear (see the next paragraph for more discussions) \cite{Xu2014}.
Each of these localized wavefunctions satisfies the symmetry of $u_{\sigma}\left(x\right)=\pm\nu_{\sigma}^{*}\left(x\right)$,
which is precisely the required symmetry for Majorana fermions. As
a result, the approximate location $\sqrt{\left\langle x^{2}\right\rangle }$
of the four Majorana fermions is exactly the same (i.e., ill-defined),
as shown in Fig \ref{fig3}(c). This superposition is very robust
with respect to the position of the dark soliton, as can be seen in
Fig. \ref{fig4}(c). When we displace the dark soliton away from the
origin to the left ($x_{1}=-0.1x_{F}$), the superposition remains,
but the relative weights of the Majorana wavefunctions localized at
the two edges are no longer equal.

It is important to note that, despite the superposition in wavefunctions,
the energies of all the Majorana fermions are nearly zero (i.e., about
$10^{-10}E_{F}$ in our numerical calculations) \cite{Xu2014}. This
fact is consistent with the earlier observation that the only one
Majorana fermion wave-function at the point node of the soliton is
superposed with the two edge Majorana fermions, so that the energy
splitting is exponentially small as in Eq. (\ref{eq:energysplitting}).
Otherwise, if the two solitonic Majorana fermions near the point node
do interfere with each other, we would rather anticipate a sizable
energy splitting $E/E_{F}\sim e^{-L/\xi_{M}}\sim1$, since now the
distance $L\sim0$.

\begin{figure}
\begin{centering}
\includegraphics[clip,width=0.48\textwidth]{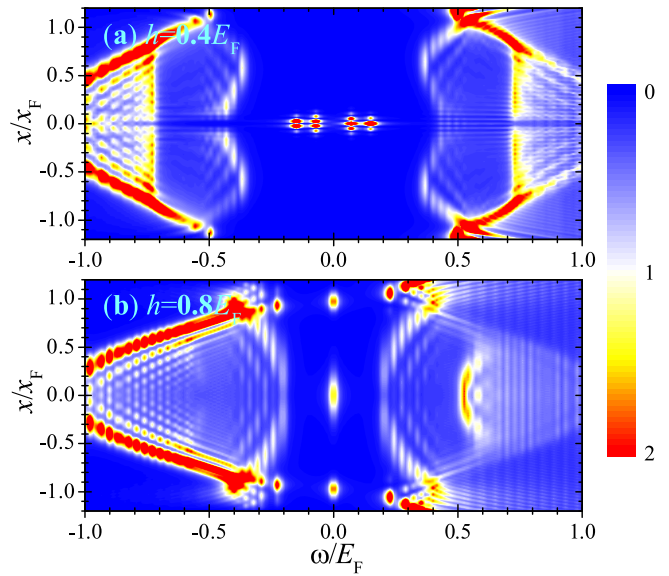} 
\par\end{centering}

\protect\caption{(color online) Local density of states of a BCS superfluid at $h=0.4E_{F}$
(a) and of a topologically-nontrivial superfluid at $h=0.8E_{F}$
(b), when a single dark soliton is created at $x_{1}=0$. In the topological
phase, the two Majorana fermions near the point node of the dark soliton
(at $x_{1}=0$) are not distinguishable because of the superposition
of the wavefunctions. The color map indicates the magnitude of the
local density of states in units of $n_{F}/E_{F}$. }

\label{fig5} 
\end{figure}

\subsection{Experimental detection of Majorana fermions}

We now turn to consider the experimental observation of the two additional
Majorana fermions localized at the point node of the dark soliton.
A \emph{direct} and convenient way is to use spatially-resolved radio-frequency
(rf) spectroscopy, which acts as a cold-atom analog of scanning tunneling
microscopy (STM) and measures the local density of states \cite{Shin2007,Jiang2011}.
In Fig. \ref{fig5}, we show the local density of states $\rho(x,\omega)$
before and after the topological transition. In the BCS superfluid
phase (a), the solitonic Andreev bound states can be easily identified.
In the topologically-nontrivial phase (b), we can clearly see the
two zero-energy Majorana fermions residing at the two trap edges.
In addition, there is a zero-energy response near the origin, arising
from the two Majorana fermions localized near the dark soliton. However,
due to their superposition (i.e., overlapping wave functions), they
are not individually resolvable.

\begin{figure}
\begin{centering}
\includegraphics[clip,width=0.48\textwidth]{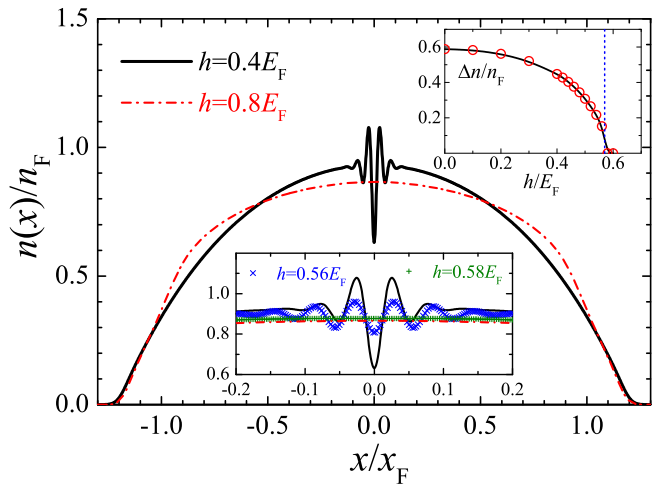} 
\par\end{centering}

\protect\caption{(color online) The total density profile in the presence of a single
dark soliton at $x_{1}=0$. The solid and dashed lines show the results
at $h=0.4E_{F}$ and $h=0.8E_{F}$, respectively. The inset at the
bottom is an enlarged view around the origin $x=0$, with additional
results near the topological transition at $h=0.56E_{F}$ (crosses)
and at $h=0.58E_{F}$ (daggers). The inset at the top right top shows
the oscillation amplitude $\Delta n$ of the density profile near
the origin, $\Delta n=n_{\textrm{max}}-n_{\textrm{min}}$, where $n_{\textrm{max}}$
and $n_{\textrm{min}}$ are the maximum and minimum densities, respectively.}

\label{fig6} 
\end{figure}

Alternatively, the existence of the soliton-induced Majorana fermions
may be \emph{indirectly} deduced from the total density profile, which
could be measured via in-situ or time-of-flight absorption imaging.
In Fig. \ref{fig6}, we present the density profile at two Zeeman
field-strengths. Before the topological transition at $h=0.4E_{F}$
(solid line), there is an apparent oscillation with spatial period
$\sim k_{F}^{-1}$, in accord with the Friedel oscillation in the
solitonic order parameter (see Fig. \ref{fig2}). As we approach the
topological transition point at the critical field strength $h_{c}\simeq0.57E_{F}$,
the amplitude of the density oscillation $\Delta n=n_{\textrm{max}}-n_{\textrm{min}}$,
where $n_{\textrm{max}}$ and $n_{\textrm{min}}$ are respectively
the maximum and minimum densities near the soliton, rapidly decreases
and vanishes precisely at the transition (see the top right inset
of Fig. \ref{fig6}). After the topological transition (see, for example,
the dashed line at $h=0.8E_{F}$ in the main figure), the density
profile becomes flat and the peak density at the trap center is essentially
independent of the Zeeman field. The disappearance of the density
oscillation is associated with the formation of the solitonic Majorana
fermion modes, whose occupation significantly contributes to the total
density because of the large amplitude of their localized wavefunctions
(see Fig. \ref{fig4}(b)). This is very similar to what happens in
the vortex core of a topological superfluid \cite{Liu2012a}. There
the core density is also greatly affected by the formation and occupation
of the vortex-core Majorana fermion modes. 

In connection with current experiments, we may consider a spin-orbit
coupled $^{40}$K Fermi gas in the presence of a tight 2D optical
lattice, with an axial trapping frequency $\omega=2\pi\times116$
Hz \cite{Wang2012}. For the typical number of total atoms $N=60$
in each tube of a 1D Fermi cloud, the Fermi temperature $T_{F}$ is
about $200$ nK. We may take $k_{R}\simeq0.75k_{F}$ and then the
recoil energy is $E_{R}\simeq0.56E_{F}$. The topological phase transition
at low temperatures (i.e., $T<0.1T_{F}\simeq20$ nK) takes place at
the critical Zeeman field $h_{c}\simeq0.6E_{F}\simeq E_{R}$, which
corresponds to a critical Rabi frequency $\Omega_{R}\simeq2E_{R}$.
The size of the Fermi cloud is about $x_{F}\simeq12$ $\mu m$. To
resolve the zero-energy Majorana modes by rf spectroscopy, we may
require the frequency/energy resolution to be better than $2\pi\times100$
Hz. On the other hand, for the in-situ density profile, the spatial
resolution needed to measure the amplitude of the density oscillations
is about $0.5\mu m$. Thus, it seems more practical to use absorption
imaging after some time-of-flight, if we assume that the structure
of density oscillations survives for a short expansion time.

\begin{figure}
\begin{centering}
\includegraphics[clip,width=0.48\textwidth]{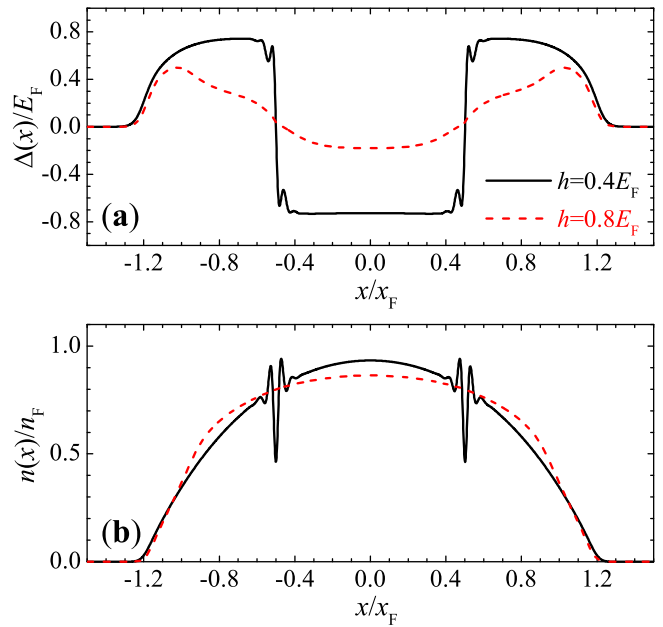} 
\par\end{centering}

\protect\caption{(color online) The solitonic order parameter (a) and the total density
distribution (b) in the presence of two dark solitons (placed at $x_{1}=-0.5x_{F}$
and $x_{2}=+0.5x_{F}$), at two Zeeman fields: $h=0.4E_{F}$ (solid
line) and $h=0.8E_{F}$ (dashed line). }

\label{fig7} 
\end{figure}

\begin{figure*}
\begin{centering}
\includegraphics[clip,width=0.98\textwidth]{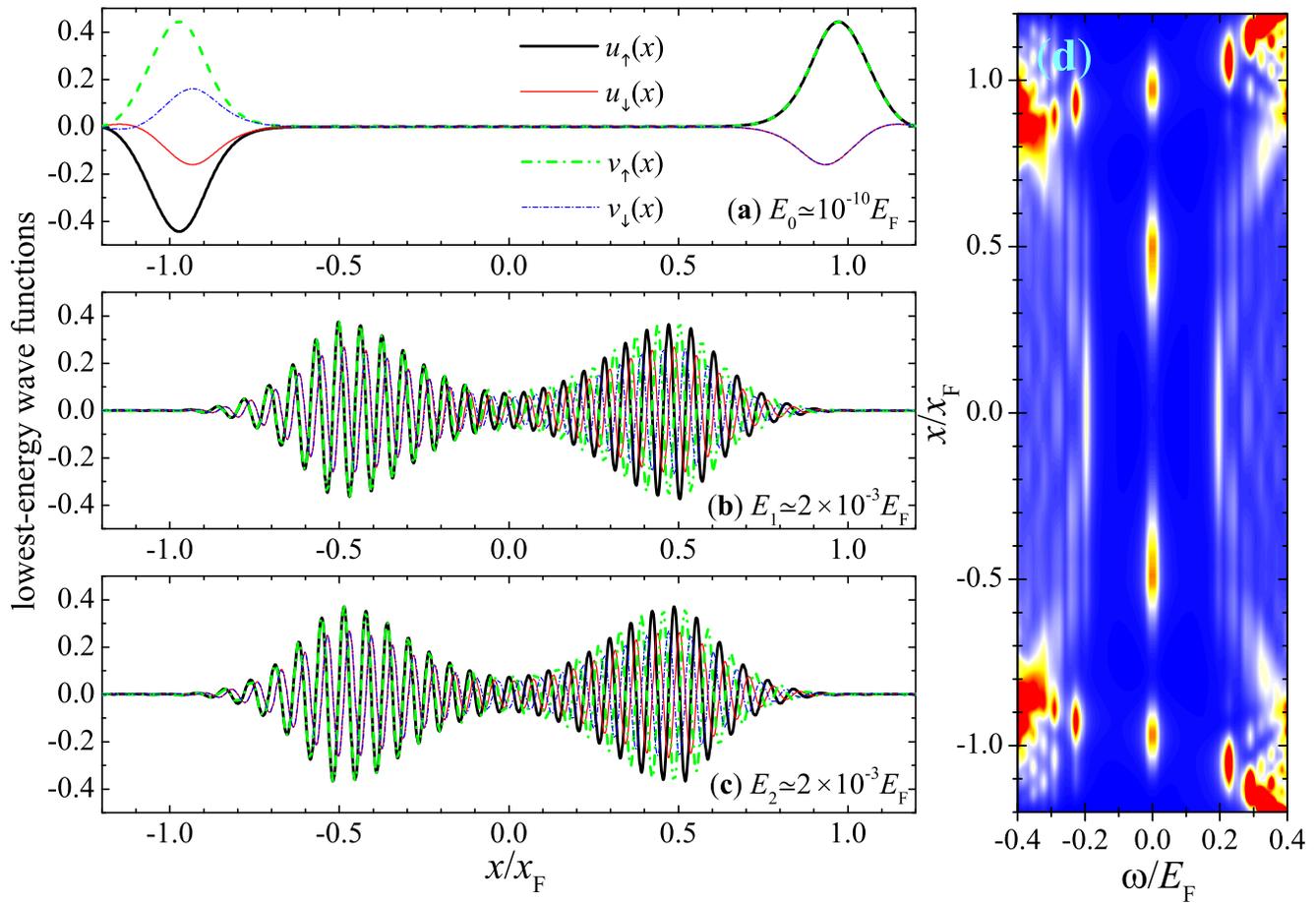} 
\par\end{centering}

\protect\caption{(color online) (a, b and c) The wave-functions of three Majorana fermions
of a topological superfluid at $h=0.8E_{F}$, in the presence of two
dark solitons at $x_{1}=-0.5x_{F}$ and $x_{2}=+0.5x_{F}$. The energy
of each Majorana fermion is indicated. (d) The corresponding local
density of state within the bulk energy gap. The color indicates the
magnitude of the local density of states, see, for example, Fig. \ref{fig5}.}

\label{fig8} 
\end{figure*}

\section{Multiple solitons}

We now consider a soliton train. Without loss of generality, let us
focus on the case of two dark solitons. Other cases with three or
more dark solitons were also examined and so far we have not found
additional, qualitatively different results. 

In Fig. \ref{fig7}, we plot the solitonic order parameter and the
total density distribution, when the two solitons are placed at $x_{1}=-0.5x_{F}$
and $x_{2}=+0.5x_{F}$. Qualitative features such as the Friedel oscillations
in the order parameter and density profile near the point nodes of
the solitons in the BCS superfluid ($h=0.4E_{F}$, solid line), can
be easily understood analogously to the case of a single dark soliton,
as discussed earlier (cf. Figs. \ref{fig2} and \ref{fig6}). 

In Fig. \ref{fig8}, we present the wavefunctions of three Majorana
fermions in the topological phase (a, b and c) and their manifestation
in the spatially-resolved rf spectrum (d). In total, there are six
Majorana fermions, localized pair-wise at the trap edges and at the
point nodes of the solitons. Only three of them are shown, one out
of each pair, owing to the particle-hole symmetry. Unlike in the case
of a single soliton, the two Majorana edge modes do not interfere
with the solitonic Majorana fermions and hence have essentially zero
energy (i.e., $E\sim10^{-10}E_{F}$) due to the exponentially small
overlap in their wavefunctions. In contrast, the overlap of the two
solitonic Majorana fermion wavefunctions - originating from the two
solitons - is significant. This leads to a nonzero energy for each
solitonic Majorana fermion, which, following Eq. (\ref{eq:energysplitting}),
is of the order of 
\begin{equation}
\frac{E}{E_{F}}\sim\exp\left(-\frac{L}{\xi_{M}}\right)\sim10^{-3}.
\end{equation}
Here, in the estimation, we have used the distance $L\sim x_{F}$
and a bit larger localization length scale $\xi_{M}\sim0.2x_{F}$
for the two solitonic Majorana wave-functions shown in Figs. \ref{fig8}(b)
and \ref{fig8}(c).

\section{Conclusions}

To summarize, we have investigated the behavior of dark solitons in
a one-dimensional topological superfluid, in the context of ultracold
atomic Fermi gases with spin-orbit coupling and an external Zeeman
field \cite{Wang2012,Liu2012b}. We have predicted that each dark
soliton can host two Majorana fermions localized at its point node,
which are detectable through the techniques of spatially-resolved
radio-frequency spectroscopy or absorption imaging. Therefore, the
well-known technique of creating dark solitons via phase imprinting
also allows one to create solitonic Majorana fermions. These Majorana
fermions can then find realistic applications in, e.g., topological
quantum information processing and quantum computation. This scheme
is very similar to the idea of using a vortex lattice in a two-dimensional
topological superfluid, where the Majorana fermions at the vortex
cores are used as qubits \cite{Ivanov2001}. In current cold-atom
experiments, it would be much easier to engineer a soliton train than
to create a vortex lattice. For the purpose of exchanging solitons
at different positions to demonstrate the non-Abelian statistics of
Majorana fermions, in future studies it would be interesting to understand
traveling grey solitons characterized by a complex order parameter
and nonzero velocity \cite{Liao2011,Scott2011,Spuntarelli2011}.

\section*{Acknowledgments}

We are grateful to Hui Hu for many helpful discussions. This work
was supported by the ARC Discovery Projects (Grant Nos. FT140100003
and DP140100637) and the National Key Basic Research Special Foundation
of China (NKBRSFC-China) (Grant No. 2011CB921502).


\begin{thebibliography}{10}
\bibitem{Frantzeskakis2010} For a review, see D. J. Frantzeskakis,
J. Phys. A \textbf{43}, 213001 (2010), and references therein.

\bibitem{Burger1999}S. Burger, K. Bongs, S. Dettmer, W. Ertmer, K.
Sengstock, A. Sanpera, G. V. Shlyapnikov, and M. Lewenstein, Phys.
Rev. Lett. \textbf{83}, 5198 (1999).

\bibitem{Denschlag2000}J. Denschlag, J. E. Simsarian, D. L. Feder,
C. W. Clark, L. A. Collins, J. Cubizolles, L. Deng, E. W. Hagley,
K. Helmerson, W. P. Reinhardt, S. L. Rolston, B. I. Schneider, and
W. D. Phillips, Science \textbf{287}, 97 (2000).

\bibitem{Dziarmaga2005}J. Dziarmaga and K. Sacha, Laser Phys. \textbf{15},
674 (2005).

\bibitem{Antezza2007}M. Antezza, F. Dalfovo, L. P. Pitaevskii, and
S. Stringari, Phys. Rev. A \textbf{76}, 043610 (2007).

\bibitem{Liao2011}R. Liao and J. Brand, Phys. Rev. A \textbf{83},
041604 (2011).

\bibitem{Scott2011}R. G. Scott, F. Dalfovo, L. P. Pitaevskii, and
S. Stringari, Phys. Rev. Lett. \textbf{106}, 185301 (2011).

\bibitem{Spuntarelli2011}A. Spuntarelli, L. D. Carr, P. Pieri, and
G. C. Strinati, New. J. Phys. \textbf{13}, 035010 (2011).

\bibitem{Cetoli2013}A. Cetoli, J. Brand, R. G. Scott, F. Dalfovo,
and L. P. Pitaevskii, Phys. Rev. A \textbf{88}, 043639 (2013).

\bibitem{Caroli1964}C. Caroli, P. G. de Gennes, and J. Matricon,
Phys. Lett. \textbf{9}, 307 (1964).

\bibitem{Yefsah2013}T. Yefsah, A. T. Sommer, M. J. H. Ku, L. W. Cheuk,
W. Ji, W. S. Bakr, and M. W. Zwierlein, Nature (London) \textbf{499},
426 (2013).

\bibitem{Ku2014}M. J. H. Ku, W. Ji, B. Mukherjee, E. Guardado-Sanchez,
L. W. Cheuk, T. Yefsah, and M. W. Zwierlein, Phys. Rev. Lett. \textbf{113},
065301 (2014).

\bibitem{Qi2011} X.-L. Qi and S.-C. Zhang, Rev. Mod. Phys. \textbf{83},
1057 (2011).

\bibitem{Wang2012} P. Wang, Z.-Q. Yu, Z. Fu, J. Miao, L. Huang, S.
Chai, H. Zhai and J. Zhang, Phys. Rev. Lett. \textbf{109}, 095301
(2012).

\bibitem{Cheuk2012} L. W. Cheuk, A. T. Sommer, Z. Hadzibabic, T.
Yefsah, W. S. Bakr, and M. W. Zwierlein, Phys. Rev. Lett. \textbf{109},
095302 (2012).

\bibitem{Williams2013}R. A. Williams, M. C. Beeler, L. J. LeBlanc,
K. Jimenez-Garcia, and I. B. Spielman, Phys. Rev. Lett. \textbf{111},
095301 (2013).

\bibitem{Majorana1937} E. Majorana, Nuovo Cimennto \textbf{14}, 171
(1937).

\bibitem{Wilczek2009} F. Wilczek, Nat. Phys. \textbf{5}, 614 (2009).

\bibitem{Liu2012a}X.-J. Liu, L. Jiang, H. Pu, and H. Hu, Phys. Rev.
A \textbf{85}, 021603(R) (2012).

\bibitem{Nayak2008} C. Nayak, S. Simon, A. Stern, M. Freedman, and
S. Das Sarma, Rev. Mod. Phys. \textbf{80}, 1083 (2008).

\bibitem{Ivanov2001}D. A. Ivanov, Phys. Rev. Lett. \textbf{86}, 268
(2001).

\bibitem{Xu2014}Y. Xu, L. Mao, B. Wu, and C. Zhang, Phys. Rev. Lett.
\textbf{113}, 130404 (2014).

\bibitem{Liu2012b} X.-J. Liu and H. Hu, Phys. Rev. A \textbf{85},
033622 (2012).

\bibitem{Wei2012} R. Wei and E. J. Mueller, Phys. Rev. A. \textbf{86},
063604 (2012).

\bibitem{Liu2013}X.-J. Liu, Phys. Rev. A \textbf{87}, 013622 (2013).
Note that, the spin-orbit coupling strength in this work should be
$\lambda=1.5E_{F}/k_{F}$, instead of $1.0E_{F}/k_{F}$ as mentioned
in the text.

\bibitem{Jiang2014}L. Jiang, E. Tiesinga, X.-J. Liu, H. Hu, and H.
Pu, Phys. Rev. A \textbf{90}, 053606 (2014).

\bibitem{Bergeman2003} T. Bergeman, M. G. Moore, and M. Olshanii,
Phys. Rev. Lett. \textbf{91}, 163201 (2003).

\bibitem{Hu2007a} H. Hu, X.-J. Liu, and P. D. Drummond, Phys. Rev.
Lett. \textbf{98}, 070403 (2007).

\bibitem{Liu2007} X.-J. Liu, H. Hu, and P. D. Drummond, Phys. Rev.
A \textbf{76}, 043605 (2007).

\bibitem{Liu2008} X.-J. Liu, H. Hu, and P. D. Drummond, Phys. Rev.
A \textbf{78}, 023601 (2008).

\bibitem{Liao2010}Y.-A. Liao, A. S. C. Rittner, T. Paprotta, W. Li,
G. B. Partridge, R. G. Hulet, S. K. Baur, and E. J. Mueller, Nature
(London) \textbf{467}, 567 (2010).

\bibitem{Mermin1966}N. D. Mermin and H. Wagner, Phys. Rev. Lett.
\textbf{17}, 1133 (1966).

\bibitem{Hohenberg1967}P. C. Hohenberg, Phys. Rev. \textbf{158},
383 (1967).

\bibitem{Guan2013}For a review, see, for example, X.-W. Guan, M.
T. Batchelor, and C. Lee, Rev. Mod. Phys. 85, 1633 (2013).

\bibitem{Lutchyn2010} R. M. Lutchyn, J. D. Sau, and S. D. Sarma,
Phys. Rev. Lett. \textbf{105}, 077001 (2010).

\bibitem{Oreg2010} Y. Oreg. G. Refael, and F. von Oppen, Phys. Rev.
Lett. \textbf{105}, 177002 (2010).

\bibitem{footnote1}In a homogeneous spin-orbit coupled Fermi gas,
the critical field is given by $h_{c}=\sqrt{\mu^{2}+\Delta^{2}}$,
at which the energy gap of the system closes \cite{Lutchyn2010,Oreg2010}.
In harmonic traps with our chosen parameters, the topological transition
occurs first at the trap center. Thus, the critical field is $h_{c}=\sqrt{\mu^{2}+\Delta_{0}^{2}},$
where $\Delta_{0}$ is the pairing gap at the trap center.

\bibitem{Potter2010}A. C. Potter and P. A. Lee, Phys. Rev. Lett.
\textbf{105}, 227003 (2010).

\bibitem{Hu2007b}H. Hu, X.-J. Liu, and P. D. Drummond, Phys. Rev.
Lett. \textbf{98}, 060406 (2007).

\bibitem{footnote3}If we take into account their hole partners, these
two physical Andreev bound states may be regarded as four modes. This
interpretation is useful for understanding the four zero-energy Majorana
fermions in the presence of a single soliton, a pair emerging from
each Andreev bound state.

\bibitem{Shin2007}Y. Shin, C. H. Schunck, A. Schirotzek, and W. Ketterle,
Phys. Rev. Lett. \textbf{99}, 090403 (2007).

\bibitem{Jiang2011}L. Jiang, L. O. Baksmaty, H. Hu, Y. Chen, and
H. Pu, Phys. Rev. A \textbf{83}, 061604(R) (2011).\end{thebibliography}
\end{document}